\documentstyle[multicol,aps,prl,epsf,rotate]{revtex}

\date{\today}
\draft
\tighten
\newcommand{\be}{\begin{equation}}
\newcommand{\ee}{\end{equation}}
\newcommand{\bea}{\begin{eqnarray}}
\newcommand{\eea}{\end{eqnarray}}
\begin{document}
\def\sqr#1#2{{\vcenter{\hrule height.3pt
      \hbox{\vrule width.3pt height#2pt  \kern#1pt
         \vrule width.3pt}  \hrule height.3pt}}}
\def\square{\mathchoice{\sqr67\,}{\sqr67\,}\sqr{3}{3.5}\sqr{3}{3.5}}
\def\today{\ifcase\month\or
  January\or February\or March\or April\or May\or June\or July\or
  August\or September\or October\or November\or December\fi
  \space\number\day, \number\year}

\def\Bbb{\bf}
\topmargin=-0.2in


\newcommand{\ww}{\mbox{\tiny $\wedge$}}
\newcommand{\pp}{\partial}

\title{Black Hole Production at LHC: String Balls and Black Holes from 
pp and Lead-lead Collisions}
\author{Andrew Chamblin$^{1,2}$ and Gouranga C. Nayak$^{1}$}
\address{$^{1}$T-8, Theory Division, Los Alamos National Laboratory,
Los Alamos, NM 87545, USA \\
$^{2}$Department of Physics,
Queen Mary, University of London,
Mile End Road, London E1 4NS \\}
\maketitle

\begin{abstract}
\\
If the fundamental planck scale is near a TeV, then parton collisions with
high enough center-of-mass energy should produce black holes.  The production
rate for such black holes at LHC has been extensively studied for the case
of a proton-proton collision. In this paper, we extend this analysis to 
a lead-lead collision at LHC. We find that the cross section
for small black holes 
which may in principle be produced in such a collision is either enhanced or
suppressed, depending upon the black hole mass. For example,
for black holes with a mass around 3 TeV we find that the differential
black hole production cross section, $d\sigma/dM$, in a typical lead-lead
collision is up to 90 times larger than that for black holes 
produced in a typical proton-proton collision. We also discuss the 
cross-sections for `string ball' production in these collisions. For string 
balls of mass about 1 (2) TeV, we find that the differential production cross section
in a typical lead-lead collision may be enhanced by a factor up to 3300 (850) 
times that of a proton-proton collision at LHC. 
\end{abstract}


\begin{multicols} {2}

\section{Introduction}

It is now generally accepted that the scale of quantum gravity 
{\it could be} as low as a TeV \cite{folks}.  If this is true, then we stand on the 
threshold of an exciting revolution in our understanding of quantum gravity
and perhaps even string theory.  One of the most exciting aspects of this revolution
will be the production of black holes in particle accelerators.  These 
`brane-world' black holes will be our first window into the extra dimensions
of space predicted by string theory, and required by the several brane-world
scenarios that provide for a low energy planck scale \cite{folks}.
While the exact metrics describing black holes in brane-world scenarios are
still largely unknown, considerable work on this issue is underway \cite{large}.
Furthermore, even without the exact metrics it is of course possible to make 
estimates based on crude information.  In particular, it is well-understood that
when the mass of the black hole is greater than the planck scale, 
the gravitational
field of the brane can be neglected; furthermore, as long as the size of the 
black hole is small compared to the characteristic length scales, then
a brane-world black hole may be regarded, to very good approximation, as simply
a higher-dimensional black hole in flat space.  
Using these approximations, in a number of recent
papers people have studied the production of microscopic black holes in
proton-proton (pp) collisions and cosmic ray events 
\cite{ppbf,pp,pp1,pp2,pp3,ag,ppch,ppk,ppu,park,hof}, 
\cite{pp4,pp5,pp6}. 

The principle aim of this paper is to extend the analysis of black hole
production to include collisions involving heavy nuclei, such as lead or 
gold at very high energy. We also estimate the string ball production
cross section both at a pp and PbPb collision at LHC. String balls are
lighter than black holes and hence they should be produced
in larger amounts at LHC. 

At present RHIC (relativistic heavy-ion colliders) at BNL collide two 
gold nuclei (to produce the `quark-gluon plasma' \cite{nay})
at $\sqrt s^{NN}$ = 200 GeV, which is insufficient to create any
black hole of TeV scale mass. This is because to create a black hole of
mass $m_B$ the minimum center of mass energy of two colliding partons must
be greater than $m_B$ which is not the case at the RHIC energy. However, 
in future
LHC will collide two lead nuclei at $\sqrt s^{NN}$ = 5.5 TeV. The total
center of mass energy of this system is $5.5 \times 208$ = 1144 TeV 
which is much larger than the 14 TeV at a pp collision at LHC. Hence, it is 
expected that many more black holes of mass less than 5.5 TeV will
be produced in a Pb-Pb collision than in a pp collision. 

Intuitively, this looks straightforward:  
A proton is a collection of partons, i.e., quarks
and gluons.  At high enough energy, a pp collision may cause some of the
partons to have a high enough center-of-mass energy to form a black hole.
Lead (at rest) consists of many protons and neutrons, and 
therefore at high energy collisions this system is just a larger
aggregate of partons. Thus, we would expect to produce more small black holes
in a lead-lead collision at $\sqrt s^{NN}$ = 5.5 TeV at LHC. An interesting
feature of a PbPb collision is that the string ball production is much more
dominant than that in a pp collision.

Of course, one should address one point here: As the energy of the colliding
partons is increased, the size of the black hole which is created may become 
large and it may absorb nearby partons. For a typical
black hole formed at LHC ($M \sim$ TeV) the rate at which the black hole
can absorb nearby partons depends on the energy density of the quarks and
gluons at LHC. The typical energy density of the quarks and gluons formed in
PbPb collisions at LHC is of the order of 1000 GeV/fm$^3$ \cite{naes}. 
It follows that the absorption is not 
very large for a TeV scale black hole at LHC \cite{prep}. 
This is because the rate of evaporation for a 
TeV scale black hole is very large \cite{eva}. In PbPb collisions
at LHC the evaporation rate of a black hole is much larger than the
absorption rate, and hence a black hole will evaporate nearly
instantaneously after its formation. In order for the black hole mass to 
be stable against decay, the energy density of the 
partons has to be very much larger
than the achievable energy density of the partons at LHC. These issues will 
be addressed elsewhere \cite{prep}.

\section{Black Hole and String Ball Production in pp and PbPb Collisions}

The black hole (string ball) production cross section 
$\sigma_{BH} (\sigma_{SB})$ at high energy hadronic collisions at zero
impact parameter is given by \cite{pp}:
\bea
\frac{d \sigma_{BH}^{AB \rightarrow BH (SB) +X}(s)}{dM^2} = {\sum}_{ab}~
\int_{\tau}^1 dx_a \int_{\tau/x_a}^1 dx_b f_{a/A}(x_a, Q^2) 
\nonumber \\
f_{b/B}(x_b, Q^2) \hat{\sigma}^{ab \rightarrow BH (SB)}(\hat s) 
~\delta(\hat s -M^2).
\label{bk1}
\eea
In the above expression $x_a (x_b)$ is the longitudinal momentum fraction
of the parton inside the hadron A(B) and $\tau=\frac{M^2}{s}$, with $\sqrt s$
being the NN center of mass energy. Energy-mometum conservation implies
$\hat s =x_ax_b s=M^2$, where $M$ is the mass of the black hole or
string ball. Using the above relation we get:
\bea
\sigma_{BH}^{AB \rightarrow BH (SB) +X}(s)={\sum}_{ab}~\frac{1}{s}
\int_{M^2>M_{min}^2} dM^2  \int_{\tau}^1 \frac{dx_a}{x_a} \nonumber \\
f_{a/A}(x_a, Q^2) 
f_{b/B}(\tau/x_a, Q^2) \hat{\sigma}^{ab \rightarrow BH (SB)}(\hat s).
\label{bk2}
\eea
$Q^2=M^2$ is the scale at which the parton distribution
function is measured. $M_{min}$ is the minimum mass of the 
black hole (string ball) above which the total cross section
is computed. ${\sum}_{ab}$ represents sum over all partonic 
combinations \cite{pp1}.
The black hole (string ball) production cross sections
in a binary partonic collision are given by \cite{pp,dp}
\bea
\hat{\sigma}^{ab \rightarrow BH }(\hat s) = \frac{1}{M^2_P}
[\frac{M_{BH}}{M_P}(\frac{8\Gamma(\frac{n+3}{2})}{n+2})]^{2/(n+1)}  ~\nonumber \\
\hat{\sigma}^{ab \rightarrow SB }(\hat s) = \frac{1}{M^2_s} ~~~~~~{\rm for}~~~~
~M_s/g_s < M_{SB} < M_s/g_s^2 \nonumber \\
\hat{\sigma}^{ab \rightarrow SB }(\hat s) = \frac{g_s^2M^2_{SB}}{M^4_s} 
~~~~~{\rm for}~~~ ~M_s < M_{SB} < M_s/g_s 
\label{bk3}
\eea
where $g_s$ is the string coupling strength, and $M_s,M_P,M_{SB}$ and $M_{BH}$ 
are the string mass scale, Planck mass scale, string ball mass and black hole
mass respectively.  n denotes the number of extra spatial dimensions.

For nuclear collisions at very high energy the calculation is similar to that
of the minijet production in AA collisions at RHIC and LHC \cite{kj}. 
The parton distribution function inside a large nucleus is given by: 
\be
R_{a/A}(x_a, Q^2) = \frac{f_{a/A}(x_a,Q^2)}{A f_{a/N}(x_a,Q^2)}
\ee
where 
$f_{a/A}(x_a,Q^2)$ and $f_{a/N}(x_a,Q^2)$ are the parton distribution functions
inside the free nucleus and free nucleon respectively. The NMC and EMC
experiments show that $R_{a/A}(x_a,Q^2) \ne 1 $ for all values of $x$.
In fact there is a strong shadowing effect ($R_{a/A}(x_a,Q^2) < 1$)
for much smaller values of $x$ ($x << 0.01$). However, for black hole
(string ball) production the shadowing effects should not be important. 
This is because we assume a minimum mass for the black hole (string ball)
$\sim$ 1 TeV, and we probe the minimum value of $x$ at $x_{min}
=M^2/s=\frac{1}{(5.5 \times 5.5)}$ =0.033 where there are no shadowing
effects. Also the present parametrizations \cite{esk,kuma}
for the ratio function $R_A(x,Q^2)$ do not cover the $Q^2$ range up to
1 $TeV^2$. For this reason we will consider the unshadowed parton 
distribution function $R_A(x,Q^2)=1$ in this paper, as there is no shadowing
for $x_{min}$=0.033 at the TeV mass scale domain of black hole (string ball) 
production. One also does not have to worry about the
saturation of parton distribution functions in PbPb collisions at LHC. 
This is because saturation happens at a very low value of $x$ 
(equivalent to $Q \sim$ 2 GeV at $\sqrt s^{NN}$ = 5.5 TeV PbPb 
collisions \cite{al})
which is much less than our minimum value $x_{min}$ =0.033. Hence for 
black hole (or string ball) production 
of TeV mass scale one does not need to worry about both shadowing and
saturation at LHC.

\section{Black hole and string ball production cross-sections at LHC:
A Comparison with pp Collisions}
In this section we compute the cross sections for black hole 
(string ball) production at LHC for a PbPb and pp collision. We use the
recent parton distribution function set CTEQ6M \cite{cteq6} in our 
calculation which is the more advanced version of CTEQ5 \cite{cteq5}.
The scale $Q$ at which the parton distribution is determined 
in our study lies within the allowed range for this PDF set. 
For black hole production we take $n=4$ throughout our calculation.
The dependence of cross section on $n$ is very weak \cite{pp}.
In Fig. 1a we present the differential cross section for black hole production
in a pp collision and in Fig. 1b we present the corresponding results
for a PbPb collision at LHC. Clearly, the cross section for 
black hole production at lower mass is much enhanced in a PbPb collision
over that in a pp collision: The differential cross section is approximately 
3300 (90) times larger for black holes of mass 1 (3) TeV. Of course,
the GR approximation fails at a mass scale that is typically larger
than the planck scale, since the minimum mass at which a black hole can
be treated general-relativistically is around $M_s/g_{s}^2$, and in
order to trust perturbative string theory the string coupling must
be less than 1.  Thus, it is unclear that it even makes sense to talk
about black holes with a mass scale right at the string scale of 1 TeV, unless
of course we are actually at strong coupling - a point to which we shall return
in the conclusions.

On the other hand, the cross section is much smaller for higher mass
black hole production in PbPb collisions because the maximum center of mass 
energy available in a binary parton collision is 5.5 TeV. As a pp collision 
is at 14 TeV c.m. energy, more larger mass black holes are produced at
this experiment. If one is trying to produce black holes with masses 
$\sim$ 5 TeV
then pp collisions at LHC constitute a good black hole production factory.

In string theory, it is now well-understood \cite{dp} that in the 
lower mass range it does not make sense to talk about black holes, but 
instead we encounter new and exotic objects known as `string balls'. 
Crudely, these are highly excited, long strings which decay 
through evaporation at the Hagedorn temperature.
We will present the results for the string ball production 
cross section using eq. (\ref{bk1}) along with the second and third expressions
of eq. (\ref{bk3}). In Fig. 2a(2b) we present the differential
cross sections for string ball production for pp (PbPb) colliisons at LHC.  
We have choosen $g_s$ =0.3 in our calculations. In Fig. 2 the 
upper(middle(lower)) line corresponds to $M_s$= 1(2(3)) TeV respectively.
For any other values of $g_s^2$ 
our results can be multiplied by an appropriate factor which is evident
from third expression of eq. (\ref{bk3}). It should be mentioned that 
for weak string coupling $g_s << 1$ the string ball production cross section
decreases and there is no black hole production as is evident from the
fact that more black holes are produced in the strong coupling limit. We also 
note that the differential cross section for 
string ball production is approximately
3300, 850, 90 times larger for string balls of mass 1, 2, 3 TeV respectively
in a PbPb collision than in a pp collision at LHC.
The ratio of the differential cross sections for
black hole production in a PbPb collision to that in a 
pp collision at LHC is the same as that of string ball production. 
This is plotted
in Fig. 3a as a function of string ball (black hole) mass. Finally we 
predict the total cross section from Eq. (\ref{bk2}) by doing a monte
carlo integration. We find that the cross sections are about 
14, resp. 28000 (nb),
in a pp, resp. PbPb, collision which produces at least 1 TeV black holes.
In Fig. 3b we present the total black hole production cross section
in pp and PbPb collisions at LHC above different values of minimum
black hole mass with $M_P$ = 1 TeV.
Note that the ${\sigma}^{PbPb}$ we obtain here is not simply 
equal to $A^2$ times
${\sigma}^{pp}$, since the PbPb collisions are at $\sqrt s^{NN}$ = 5.5 TeV,
whereas the pp collisions are at $\sqrt s^{NN}$ = 14 TeV.

\section{Conclusion}
If the fundamental planck scale is near a TeV, then parton collisions with
high enough center-of-mass energy should produce black holes and string balls. 
Black hole production in pp a collision at LHC
has been extensively studied in the literature. 
In this paper we have calculated the cross sections for black hole
and string ball production at LHC in a PbPb collision at 
$\sqrt s^{NN}$ = 5.5 TeV. We have also computed the string ball production
cross section for a pp collision at $\sqrt s^{NN}$ = 14 TeV at LHC.
We find that the cross section of small black holes which
may in principle be produced in such collisions is either enhanced or
suppressed, depending upon the black hole mass.  For example,
for black holes with a mass around 3 TeV we find that a typical lead-lead
collision may produce up to 90 times the $d\sigma/dM$ of such black holes 
that would be produced in a typical proton-proton collision. For string 
balls of mass about 1, 2, 3 TeV we find that a typical lead-lead
collision may produce up to 3300, 850, 90 times the $d\sigma/dM$ of such 
string balls that would be produced in a proton-proton collisions at LHC. 

The key mass scale in this story is the minimum mass, $M_{s}/g_{s}^2$, at which 
it makes
sense to talk about a black hole, and at which we see the transition to string
balls. Obviously, this mass scale depends explicitly upon the string coupling.
Furthermore, it has the intriguing property that we {\it lower} the mass scale
for black holes as we run to stronger coupling. This is particularly interesting
from the vantage point of recent ideas which are emerging in string cosmology.
In particular, it has been suggested that several problems, including even the
famous cosmological constant problem, may be resolved in scenarios where the
dilaton or string coupling {\it runs to infinity} \cite{cosmo}.  In such a 
scenario, it is clear that the string ball picture must break down, since we
are in the strongly coupled regime or `M-theory'.  Indeed, if string theory is 
strongly coupled, the window for string balls presumably
disappears, and all massive `string' states will be black holes.  This means
that it may be possible to place strong experimental contraints on these cosmological
models with a runaway dilaton, if we do {\it not} see black holes in collider 
experiments.  Research on this and related issues is currently underway.

\acknowledgments

We are grateful to Roberto Emparan for useful communications.
AC was supported by a Director's Funded Fellowship at Los Alamos National Lab,
where this research was supported by the Department of Energy, under
contract W-7405-ENG-36.

\end{multicols}

\vspace{2.5cm}

\parbox{15cm}{
\parbox[t]{7cm}
{\begin{center}
\mbox{\epsfxsize=6.5cm\epsfysize=4cm\epsfbox{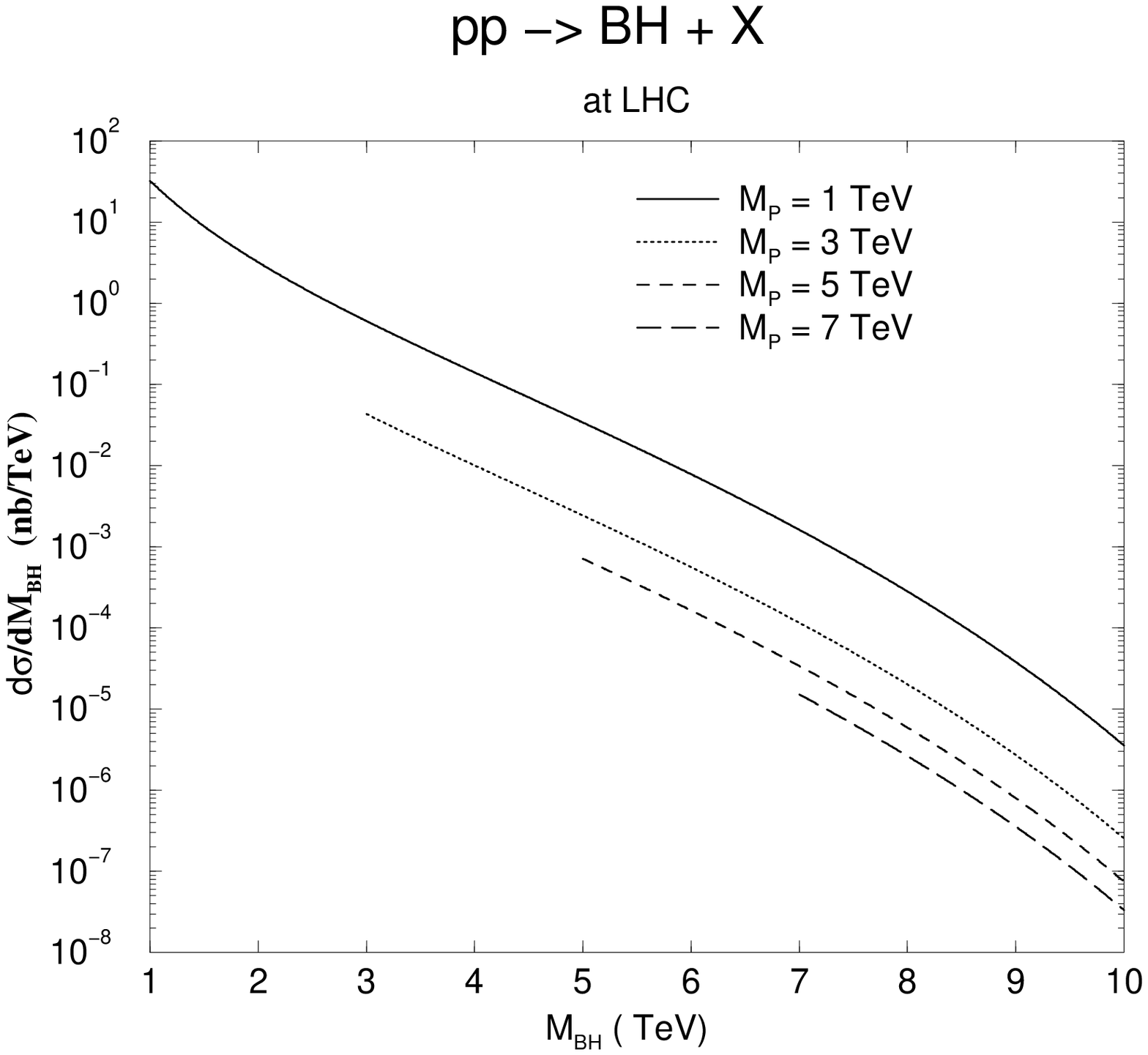}}
\end{center}}
\hspace{.5cm}\parbox[t]{7cm}
{\begin{center}
\mbox{\epsfxsize=6.5cm\epsfysize=4cm\epsfbox{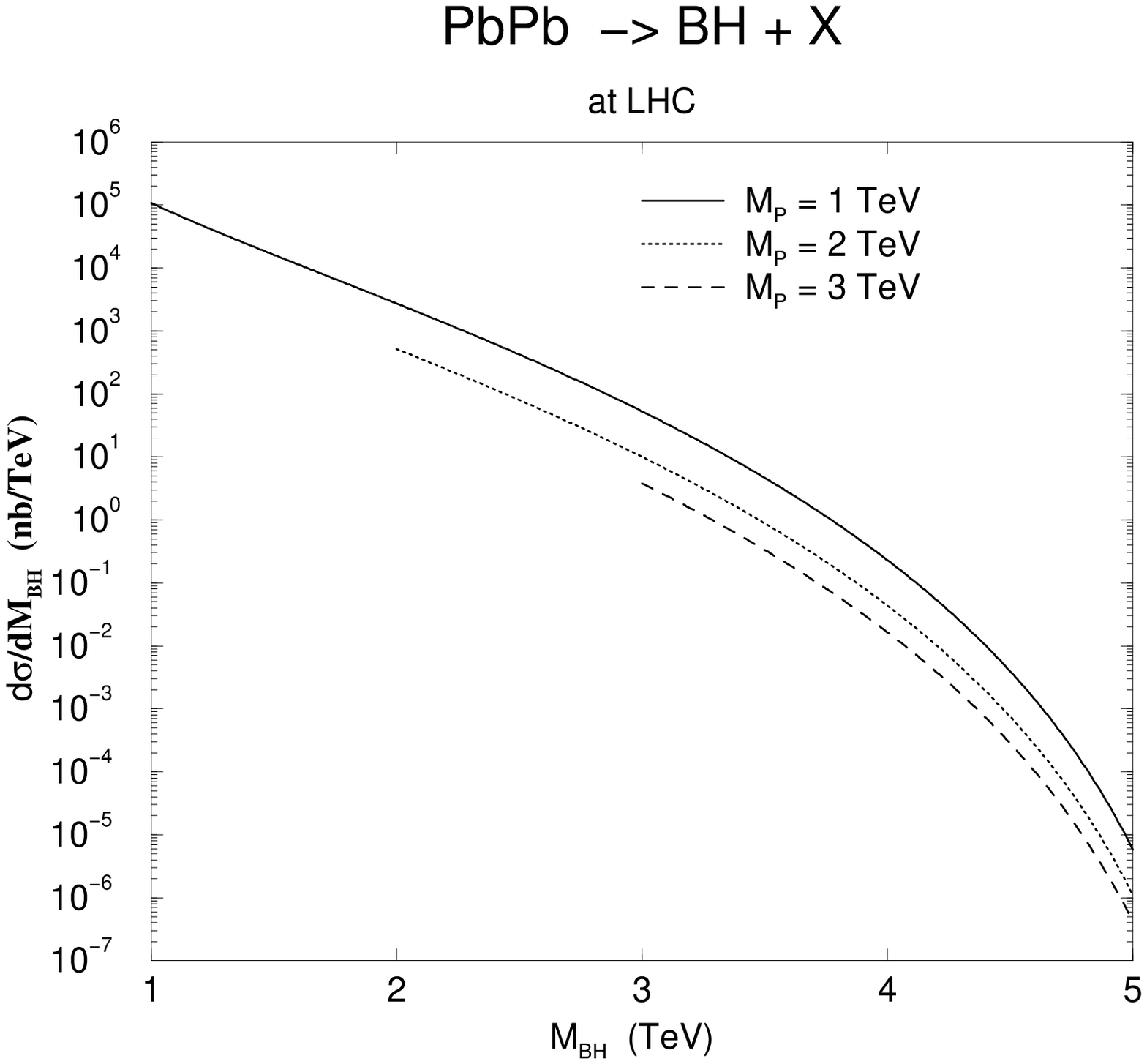}}
\end{center}}
\parbox[t]{7cm}
{{\small FIG. 1a: The differential cross section for black hole production
$d\sigma/dM_{BH}$ in a pp collision at $\sqrt s^{NN}$ = 14 TeV at LHC.
}}
\hspace{0.8cm}\parbox[t]{7cm}
{{\small FIG. 1b:
The differential cross section for black hole production $d\sigma/dM_{BH}$ in
a PbPb collision at $\sqrt s^{NN}$ = 5.5 TeV at LHC.
}}}

\vspace{2.5cm}

\parbox{15cm}{
\parbox[t]{7cm}
{\begin{center}
\mbox{\epsfxsize=6.5cm\epsfysize=4cm\epsfbox{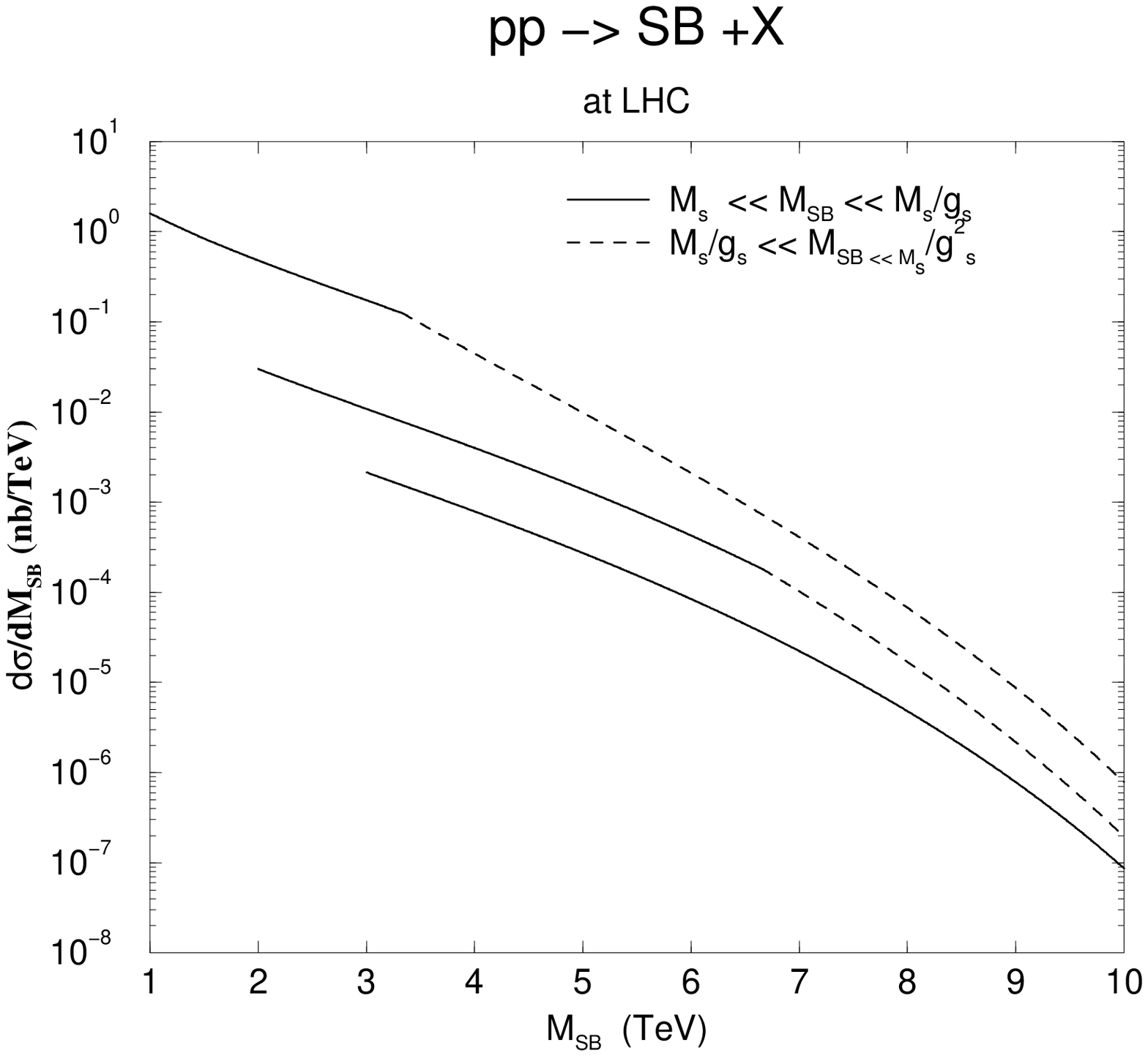}}
\end{center}}
\hspace{.5cm}\parbox[t]{7cm}
{\begin{center}
\mbox{\epsfxsize=6.5cm\epsfysize=4cm\epsfbox{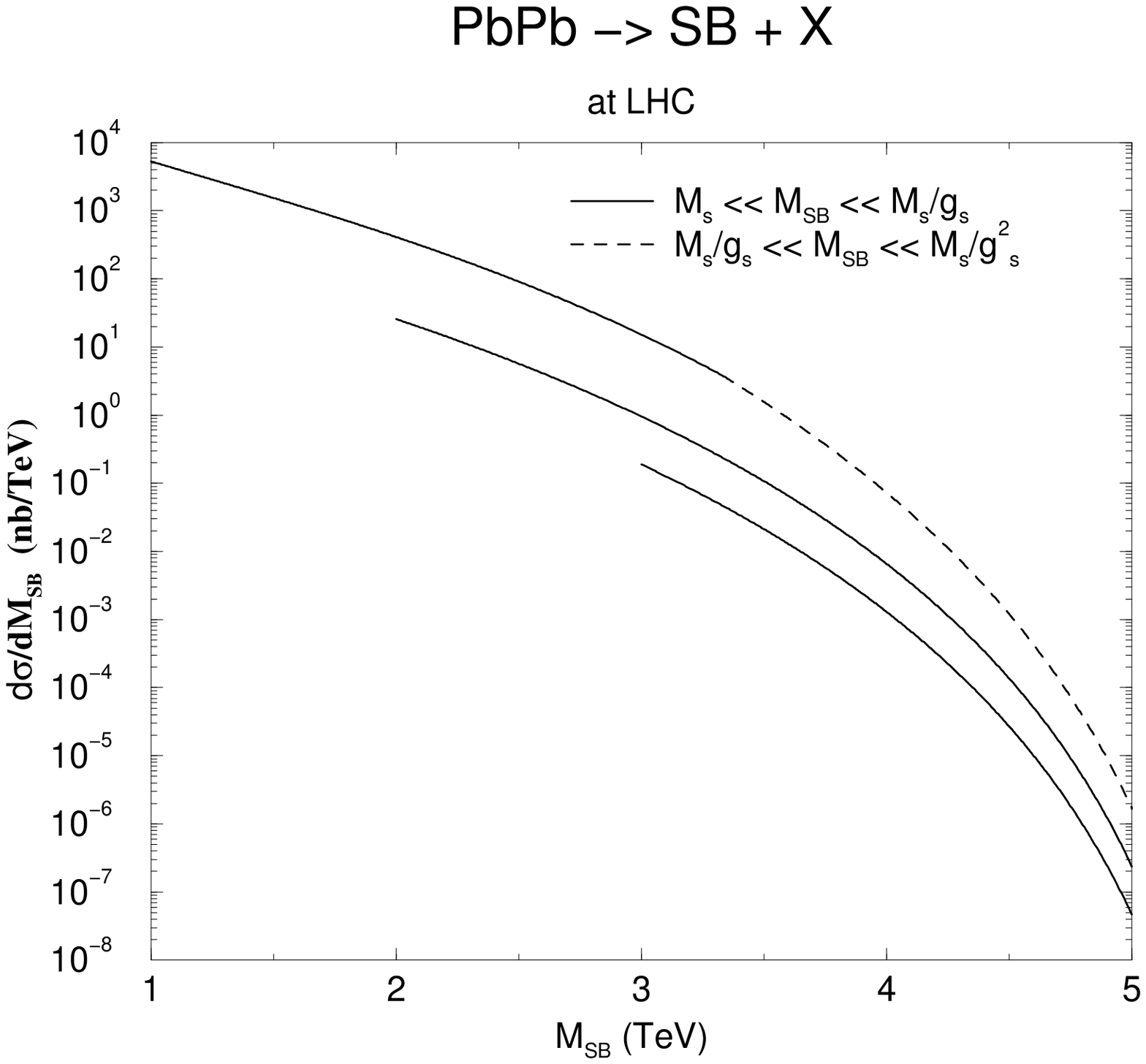}}
\end{center}}
\parbox[t]{7cm}
{{\small FIG. 2a: The differential cross section for string ball
production
$d\sigma/dM_{SB}$ in a pp collision at $\sqrt s^{NN}$ = 14 TeV at LHC.
Three curves corresponds to three different values of $M_s$ used, as
discussed in the text. $g_s$ = 0.3 used in the calculation}}
\hspace{0.8cm}\parbox[t]{7cm}
{{\small FIG. 2b:
The differential cross section for string ball 
production $d\sigma/dM_{SB}$ in a
PbPb collision at $\sqrt s^{NN}$ = 5.5 TeV at LHC. Other parameters are
same as in FIG. 2a.
}}}

\vspace{2.5cm}

\parbox{15cm}{
\parbox[t]{7cm}
{\begin{center}
\mbox{\epsfxsize=6.5cm\epsfysize=4cm\epsfbox{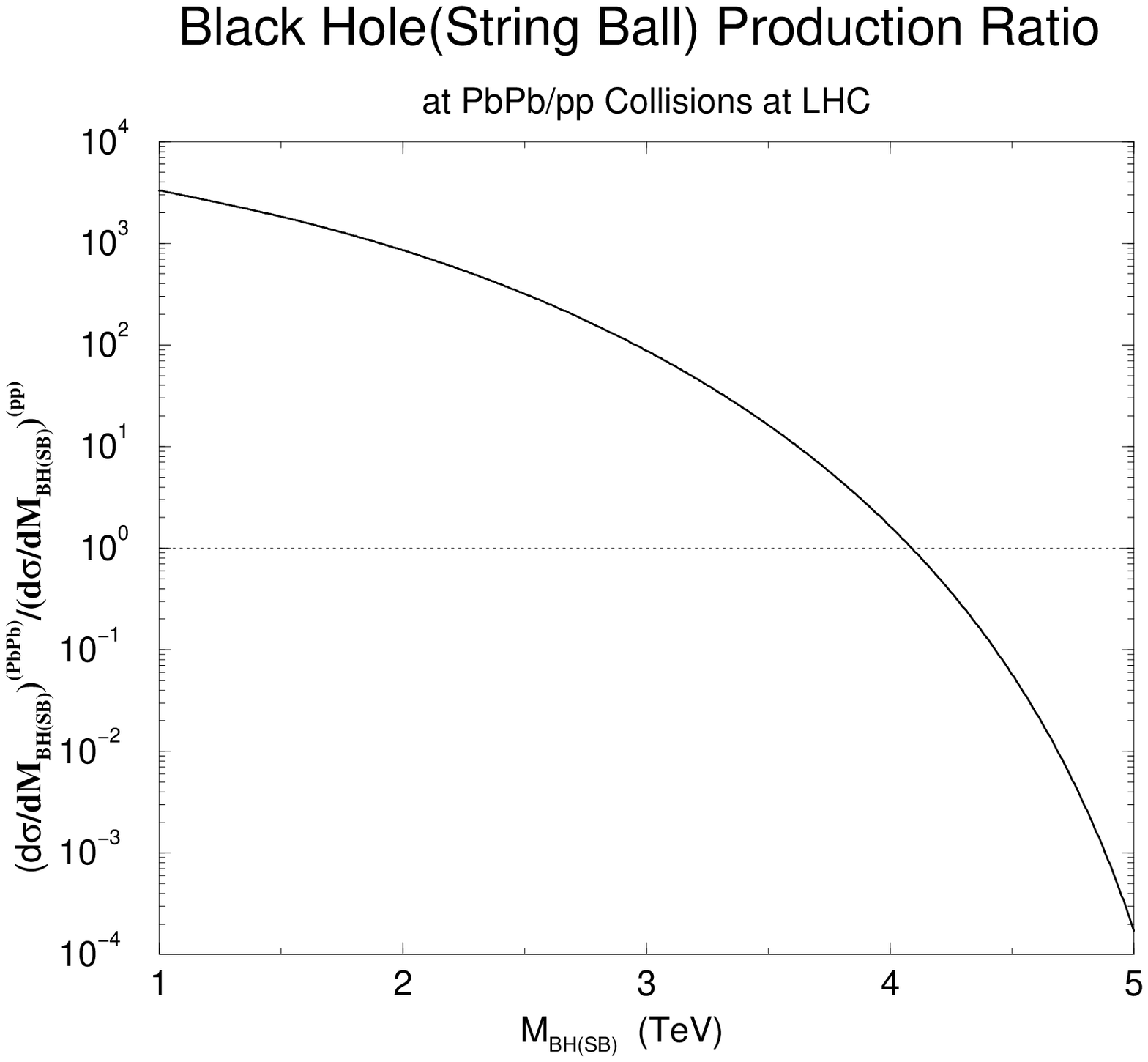}}
\end{center}}
\hspace{.5cm}\parbox[t]{7cm}
{\begin{center}
\mbox{\epsfxsize=6.5cm\epsfysize=4cm\epsfbox{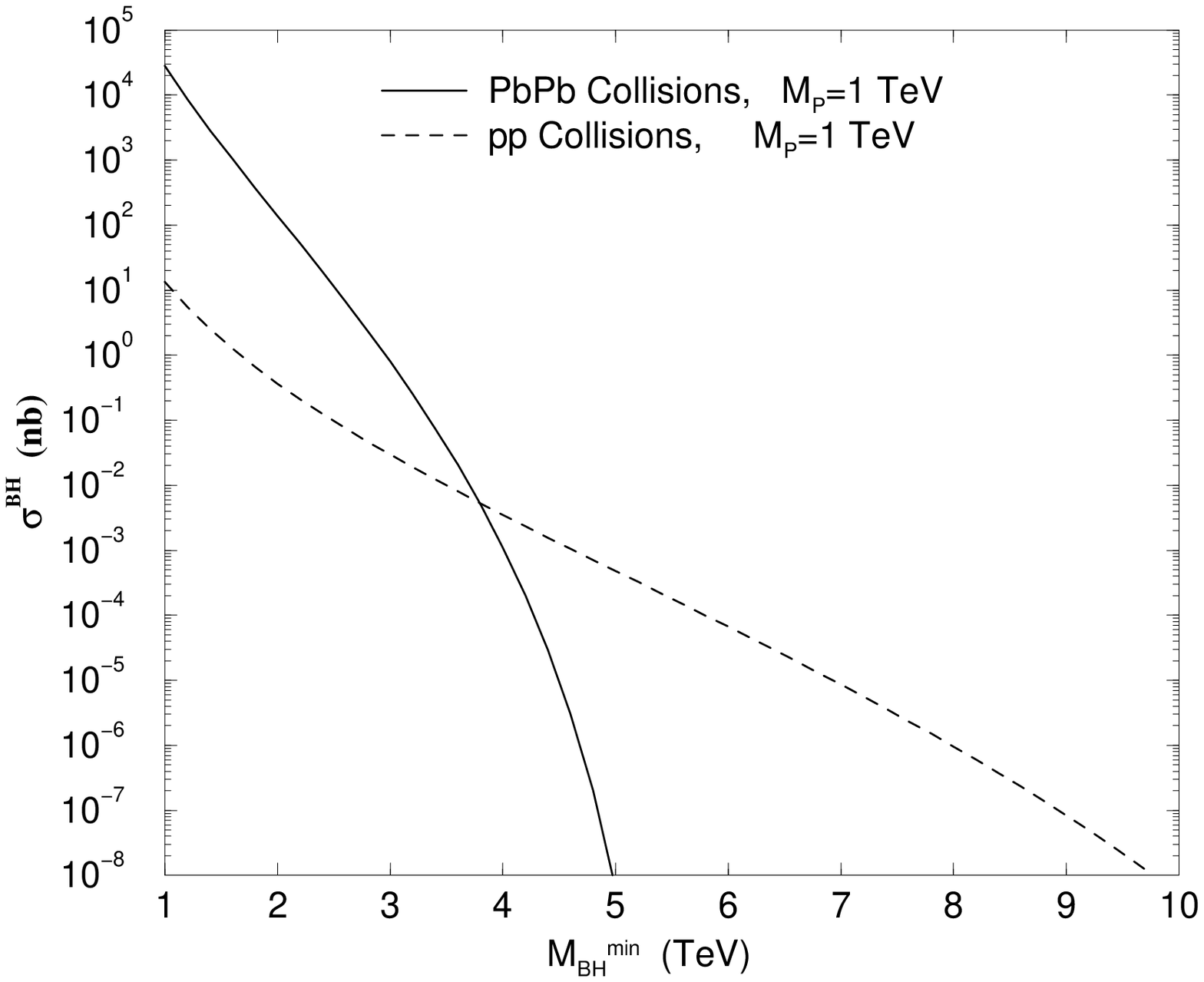}}
\end{center}}
\parbox[t]{7cm}
{{\small FIG. 3a: 
Ratio of the black hole (string ball) production in a PbPb/pp collision
at LHC as a function of black hole (string ball) mass.}}
\hspace{0.8cm}\parbox[t]{7cm}
{{\small FIG. 3b:
The total cross section for black hole 
production in a
PP collision at $\sqrt s^{NN}$ = 14 TeV at LHC and in a
PbPb collision at $\sqrt s^{NN}$ = 5.5 TeV at LHC
}}}

\end{document}